\documentclass[showpacs,showkeys,superscriptaddress,preprint]{revtex4}

\usepackage{amsmath}

\newcommand{\qvec}{\boldsymbol{q}}
\newcommand{\kvec}{\boldsymbol{k}}
\newcommand{\Kvec}{\boldsymbol{K}}
\newcommand{\pivec}{\boldsymbol{\pi}}
\newcommand{\zerovec}{\boldsymbol{0}}

\newcommand{\cdagg}{c^{\dagger}}
\newcommand{\ddagg}{d^{\dagger}}
\newcommand{\alphadagg}{\alpha^{\dagger}}
\newcommand{\betadagg}{\beta^{\dagger}}

\usepackage{graphicx}

\begin{document}

\title{Towards analytical approaches to the dynamical-cluster approximation}

\author{J.P.Hague} 

\affiliation{Department of Physics and Astronomy, The Open University, Walton Hall, Milton Keynes, MK7 6AA, UK}

\date{30th November 2009}

\begin{abstract}
I introduce several simplified schemes for the approximation of the
self-consistency condition of the dynamical cluster approximation. The
applicability of the schemes is tested numerically using the
fluctuation-exchange approximation as a cluster solver for the Hubbard
model. Thermodynamic properties are found to be practically
indistinguishable from those computed using the full self-consistent
scheme in all cases where the non-interacting partial density of
states is replaced by simplified analytic forms with matching 1st and
2nd moments. Green functions are also compared and found to be in
close agreement, and the density of states computed using Pad\'{e}
approximant analytic continuation shows that dynamical properties can
also be approximated effectively. Extensions to two-particle
properties and multiple bands are discussed. Simplified approaches to
the dynamical cluster approximation should lead to new analytic
solutions of the Hubbard and other models.
\end{abstract}


\maketitle

\section{Introduction}
\label{section:introduction}

A major effort in condensed-matter theory goes towards the development
of new techniques for the study of correlated-electron systems, such
as the dynamical mean-field theory (DMFT). When studying the Hubbard
model, DMFT is often implemented with a simple analytic form of the
self-consistency condition that corresponds to using a Gaussian,
semi-circular or Lorentzian non-interacting density of states
\cite{metzner1989a,georges1996a}. This has led to analytic work
investigating, for example the Mott transition
\cite{rozenberg1994a}. Results from dynamical mean-field theory are
expected to be applicable to 3D systems, but since DMFT only considers
local processes, physics in low dimensions cannot accurately be
investigated.

There have been several approaches that extend the core ideas of DMFT
so that low-dimensional systems can be studied, including the
dynamical-cluster approximation (DCA) which systematically
reintroduces spatial fluctuations into the dynamical mean-field theory
\cite{hettler1998a}. DCA is an extremely powerful method which has
been applied to a wide range of model systems \cite{maier2005a}, but
it can be computationally expensive, and intensive numerical
simulation may lack intuition. Carrying out analytic work with the
dynamical-cluster approximation is difficult owing to the
coarse-graining step which completes the self-consistency, which
consists of a partial sum over momenta in small regions of the
Brillouin zone. The partial sum can not be carried out analytically.

\begin{figure}
\includegraphics[width=70mm]{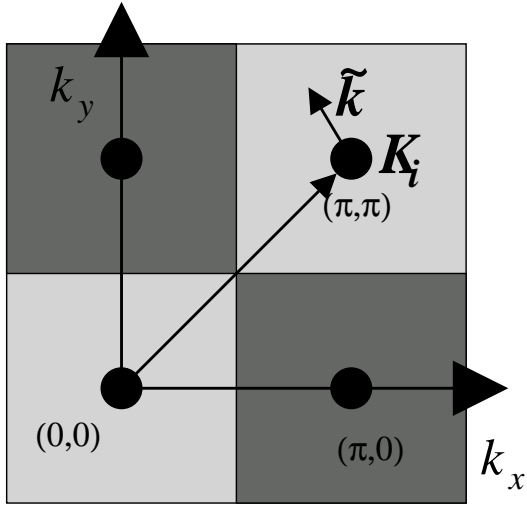}
\caption{Schematic of the dynamical cluster approximation. The
Brillouin zone is split into sub-zones, within which the self-energy
is assumed to be constant. The centers of the zones are at the momenta
$\Kvec_i$ and a vector $\tilde{\kvec}$ is defined which relates the
center of a sub-zone to a location in that sub-zone.}
\label{fig:summary}
\end{figure}

By neglecting momentum conservation at the interaction vertices, DMFT
replaces the self-energy, $\Sigma(\kvec,\omega)$, by its
momentum-independent counterpart. In 2D, the spatial fluctuations
which were neglected in the DMFT are expected to contribute
significantly. DCA includes these fluctuations by dividing up the
Brillouin zone into sub-zones which obey the lattice symmetry
\cite{hettler1998a} or more recently using Bett's clusters \cite{kent2005a}. This coarse graining step corresponds to a
mapping onto a cluster-impurity model. Within each of these sub-zones,
the self-energy is assumed to be momentum independent, so the coarse
graining of the Green function can be written
\begin{equation}
G(\Kvec_i,z)=\sum_{\tilde{\kvec}}\frac{1}{z-\epsilon(\Kvec_i+\tilde{\kvec})-\Sigma(\Kvec_i,z)} \, .
\end{equation}
Here $\tilde{\kvec}$ is a vector that lies within the sub-zone as shown in Fig. \ref{fig:summary}. The sum over momenta is normally carried out numerically.

Since the self-energy is momentum independent within each sub-zone, it is possible to replace the sum over momenta with a partial density of states, leading to the relation \cite{hettler1998a},
\begin{equation}
\label{eqn:dcagreensfunction}
G(\Kvec_{i},z)=\int _{-\infty }^{\infty }\frac{D_{\Kvec_i}(\epsilon )\, d\epsilon }{z-\epsilon -\Sigma (\Kvec_{i},z)}\, ,
\end{equation}
where $D_{\Kvec_i}(\epsilon )$ is the non-interacting fermion partial
density of states (DOS) for the sub-zone centered about $\Kvec_i$ (see
figure \ref{figure:comparedos}(a)).

In both DMFT and DCA, the self-energy and the coarse-grained Green
function are also related through a modified Dyson equation,
\begin{equation}
\label{eqn:moddyson}
{\mathcal{G}}_{0}^{-1}(\Kvec_{i},z)-G^{-1}(\Kvec_{i},z)=\Sigma (\Kvec_{i},z) \, .
\end{equation}
The self-consistency is closed by calculating the self-energy
from
${\mathcal{G}}_{0}(\Kvec_{i},z)$, which may be interpreted as the host Green function of an impurity
model \cite{georges1996a}. Cluster impurity problems have been extensively studied,
therefore a number of different approximations for the self-energy are
readily available.

Here, I suggest a simple analytical form for the dynamical-cluster
self-consistency step, Eqn. \ref{eqn:dcagreensfunction}. I use results
from the Hubbard model \cite{hubbard1963a} to establish if the
simplified forms lead to acceptable results. The Hubbard model is one
of the simplest non-trivial models of electronic correlations in the
solid state, and examines a single band of electrons, hopping between
lattice sites with amplitude $t$, and interacting via a Coulomb
repulsion, $U$, represented by a Hamiltonian,
\begin{equation}
H=-\sum_{<ij>\sigma}t_{ij}c^{\dagger}_{i\sigma}c_{j\sigma} + U \sum_i n_{i\uparrow}n_{i\downarrow}
\end{equation}
($c^{\dagger}_{i\sigma}$ creates an electron on site $i$ with spin $\sigma$ and $n_{i\sigma}$ is
the number operator). This model can only be solved
exactly in one dimension \cite{lieb1968a}, but has been extensively
examined in the infinite-dimensional limit corresponding to DMFT
\cite{georges1996a}, and in a wide range of other approximations,
including DCA \cite{maier2005a}.

This paper is set out as follows: In section \ref{section:dca}, I
suggest and summarize several approximate forms for the
self-consistency step of the DCA equations. I test the accuracy of the
new schemes in section \ref{section:results}. Possible extension for
two-particle properties is discussed in section \ref{sec:extension}
and for multiple bands in section \ref{section:multiband}. A brief
discussion can be found in section \ref{section:summary}.

\section{Analytical approximations to DCA}
\label{section:dca}

\begin{table}
\begin{ruledtabular}
\caption{Offsets and widths of the partial DOS for a 2D tight-binding
square lattice to be used as input for the various schemes. First and
second moments of the irreducible partial densities of states are used
to calculate these values up to cluster sizes of $N_c=8$ ($W=4t$ is the half band-width).}
\begin{tabular}{ccccc}
$N_C$ & $\Kvec_i=(k_x,k_y)$ & Offset ($\bar{\epsilon}_{\Kvec_i}/W$) & Width ($t_i/W$) \\
\hline
1 & $(0,0)$ & 0.000  &  0.500 \\
\hline
2 & $(0,0)$ & 0.405 & 0.293 \\
  & $(\pi,\pi)$ & -0.405 & 0.293 \\
\hline
4 & $(0,0)$ & 0.637 & 0.218 \\
  & $(\pi,0),(0,\pi)$ & 0.000 & 0.218 \\
  & $(\pi,\pi)$ & -0.637 & 0.218 \\
\hline
8 & $(0,0)$ & 0.811 & 0.112 \\
  & $(\pi/2,\pi/2),(3\pi/2,\pi/2),$ & & \\
  & $(\pi/2,3\pi/2),(3\pi/2,3\pi/2)$ & 0.000 & 0.386 \\
  & $(\pi,\pi)$ & -0.811 & 0.112 \\
  & $(\pi,0),(0,\pi)$ & 0.000 & 0.181 \\
\end{tabular}
\end{ruledtabular}
\end{table}

An appealing aspect of DMFT is its potential for analytic work. Three
forms for the non-interacting density of states are commonly used: (1)
a Gaussian, which corresponds to a hyper-cubic lattice with high
dimensionality, (2) a semicircular density of states which relates to
a Bethe lattice with large coordination number and (3) a Lorentzian,
which decouples the self-consistent equations linking DMFT with the
Anderson impurity model. In the case of the DCA, the non-interacting
partial density of states (partial DOS) has a complicated form, which
prevents easy analytic work. In the very large cluster limit each
partial DOS tends towards a $\delta$-function form. Thus in principle,
any approximate partial DOS with the properties of a $\delta$ function
has the correct large cluster behavior (for example one might replace the
partial DOS with a Gaussian). One may go further and match the mean
($\bar{\epsilon}_{\Kvec_{i}}$) and variance
($t_i=\sqrt{\bar{\epsilon^2}_{\Kvec_{i}}-(\bar{\epsilon}_{\Kvec_{i}})^2}$)
for the approximate and exact partial DOS, to get the correct large
cluster asymptotic behavior (since the low order moments dominate the form
of the self-consistent condition in that case). Here the bar indicates
an average over the momentum states in the subzone centered at
$\Kvec_i$. Smaller cluster sizes are likely to be most useful for
analytic work of the type carried out with DMFT, thus it is of
interest to determine whether such a procedure is useful when
$N_C=4$, which is the smallest cluster size which can display low
dimensional properties in 2D.

\begin{figure}
\includegraphics[width=120mm, height=160mm]{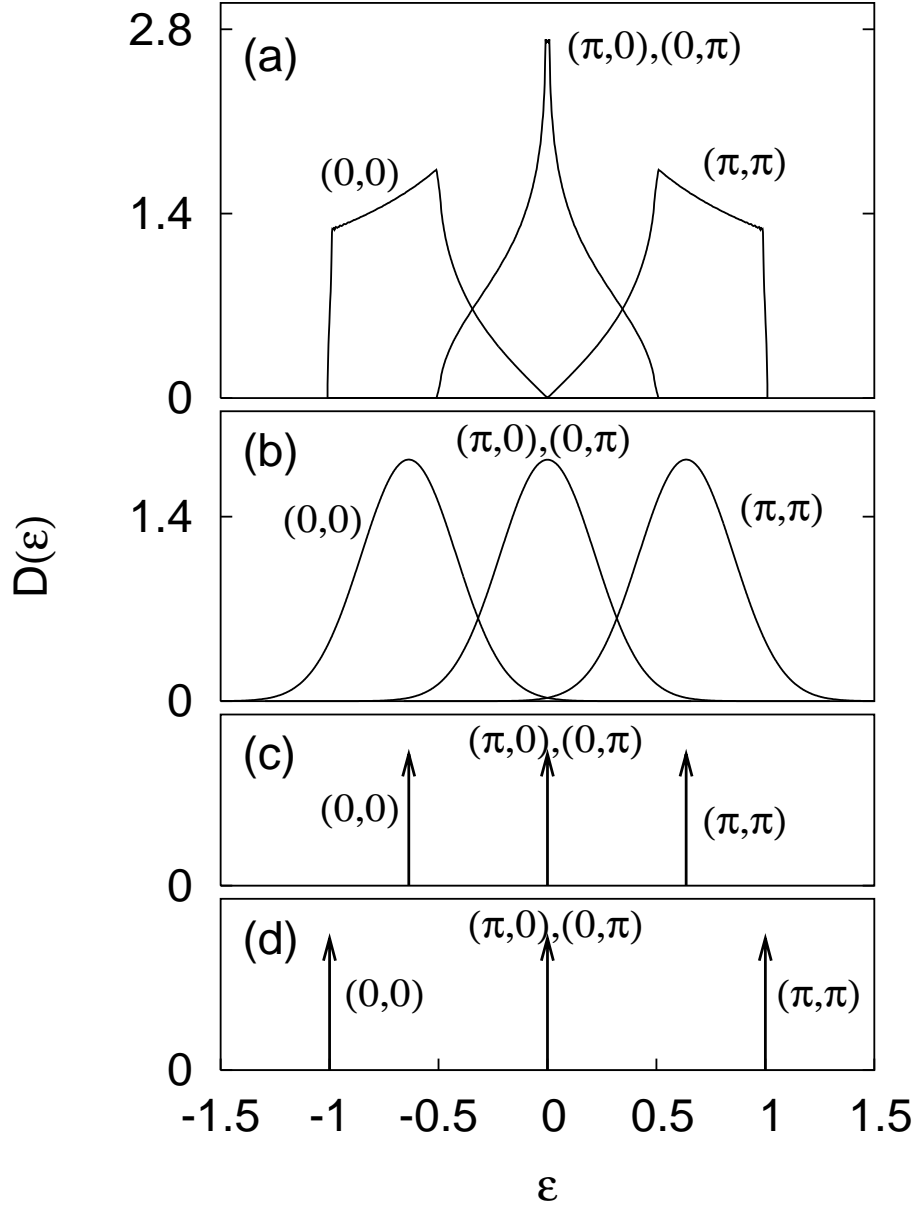}
\caption{Comparison between the partial non-interacting densities of states for
(a) the DCA (calculated using the analytic tetrahedron method \cite{lambin1984a}), (b)
Gaussian approximation to the DCA self-consistency (c) modified finite size calculations and
(d) finite size calculations. A cluster size of $N_C=4$ is shown since
this is expected to be the most useful for analytic work. As the
lattice size increases, the Gaussian form converges with the exact result.}
\label{figure:comparedos}
\end{figure}

The essence of the proposed approach is shown for a cluster size of
$N_C=4$ in figure \ref{figure:comparedos} for a quasi-2D system with
in plane hopping, $t_{\parallel}=25t_{\perp}$ (where $t_{\perp}$ is
the inter-plane hopping). Panel (a) shows the numerically exact
partial DOS for $N_C=4$.  Panel (b) shows the replacement of the
partial DOS by Gaussians which match the first 2 moments. Panel (c)
shows a modified finite size approximation, which replaces the partial
DOS with a delta function matching the 1st moment only. The equivalent
non-interacting ``densities of states'' associated with a finite size
calculation are shown in panel (d). With a simple approximate form for
the partial density of states, the cluster Green function can easily
be calculated. I propose several approximate forms for the partial DOS
for which an analytic form for the self-consistent condition can be
found. These are summarized for convenience in table
\ref{tab:schemes}, along with the analytic form of the Green
function. A key feature of the approximation considered in this
article is that it becomes better as cluster size increases since the
second moment becomes vanishingly small, and the forms of the partial
DOS converge. As such, examinations of the four-site cluster ($N_C=4$)
represent the most challenging test of the approximate scheme.

\begin{table*}
\begin{ruledtabular}
\caption{Analytic forms for the partial DOS and the corresponding
self-consistent
equation. $z_i=i\omega_n+\mu-\Sigma(\Kvec_i,i\omega_n)$,
$s=\mathrm{sgn}(\mathrm{Im}[z_i])$, $\mathrm{erfc}$ is the
complimentary error function,
$t_i=\sqrt{\bar{\epsilon}^2_{\Kvec_{i}}-(\bar{\epsilon}_{\Kvec_{i}})^2}$
and $\bar{\epsilon}_i$ is the average of the non-interacting
dispersion over the course graining zone.}
\begin{tabular}{ccc}
Scheme & partial DOS ($D_i(\epsilon)$) & Green function ($G(z_i)$)\\
\hline
Gaussian & $\frac{\exp(-(\epsilon-\bar{\epsilon}_i)^2/2t_i^2)}{t_i \sqrt{2\pi}}$ & $-\frac{is\sqrt{\pi}}{t_i\sqrt{2}}\exp\left(-\frac{(z-\bar{\epsilon}_i)^2}{2t_i^2}\right)\mathrm{erfc}(-\frac{is(z-\bar{\epsilon}_i)}{t_i\sqrt{2}})$\\
\hline
Semi-circular & $\frac{\sqrt{4t_i^2-(\epsilon-\bar{\epsilon}_i)^2}}{2\pi t_i^2}$ ($|\bar{\epsilon_i}-\epsilon| < 2t_i$)
 & $\frac{z_i-\bar{\epsilon}_i-s\sqrt{(z_i-\bar{\epsilon}_i)^2-4t_i^2}}{2t_i^2}$
 \\
\hline
Square & $\frac{1}{2t_i\sqrt{3}} \hspace{3mm}(|\epsilon-\bar{\epsilon}_i|<t_i\sqrt{3})$ & $\frac{1}{2t_i\sqrt{3}}\ln\left[\frac{z-\bar{\epsilon_i}-t_i\sqrt{3}}{z-\bar{\epsilon_i}+t_i\sqrt{3}}\right]$ \\
\hline
Lorentzian & $\frac{t_i}{\pi((\epsilon-\bar{\epsilon}_i)^2+t_i^2)}$
 & $\frac{1}{z_i-\bar{\epsilon}_i+it_i\mathrm{sgn}(\mathrm{Im}[z_i])}$
 \\
\hline
Modified FS & $\delta(\epsilon-\bar{\epsilon}_i)$
& $\frac{1}{z-\bar{\epsilon}_i}$
 \\
\end{tabular}
\end{ruledtabular}
\label{tab:schemes}
\end{table*}

The Gaussian approximation to the partial DOS has the advantage that
the full moment expansion is uniquely defined by the first two moments
in the expansion, so all moments are well defined. However, the form of the Green function is
relatively complicated, involving a complimentary error function. The
semi-circular and square partial DOS are bounded, as is the case for all
partial DOS in low dimensional systems, and lead to a Green function
with a simple analytic form (i.e. the self-consistent
equation can be inverted in terms of common functions). The first and
second moments of the partial DOS are matched in all of these
schemes. The square DOS approximation is not tested
here.

There are two additional schemes that have some potential for analytic
calculations. The first is the Lorentzian approximation. For $N_C=1$
this corresponds to the solution of an Anderson impurity model, and as
such the self-consistent equations are decoupled. The second treats the partial
DOS as a $\delta$-function, where only the 1st moment is matched. This
is similar to using a finite size (FS) technique, except that the
shift in the peak location from the original position involves a large
number of states, and such an approximation has aspects of both the
thermodynamic limit of a large number of states and of finite size simulations.

\section{Numerical test of applicability}
\label{section:results}

\begin{figure}
\includegraphics[height=90mm,angle=270]{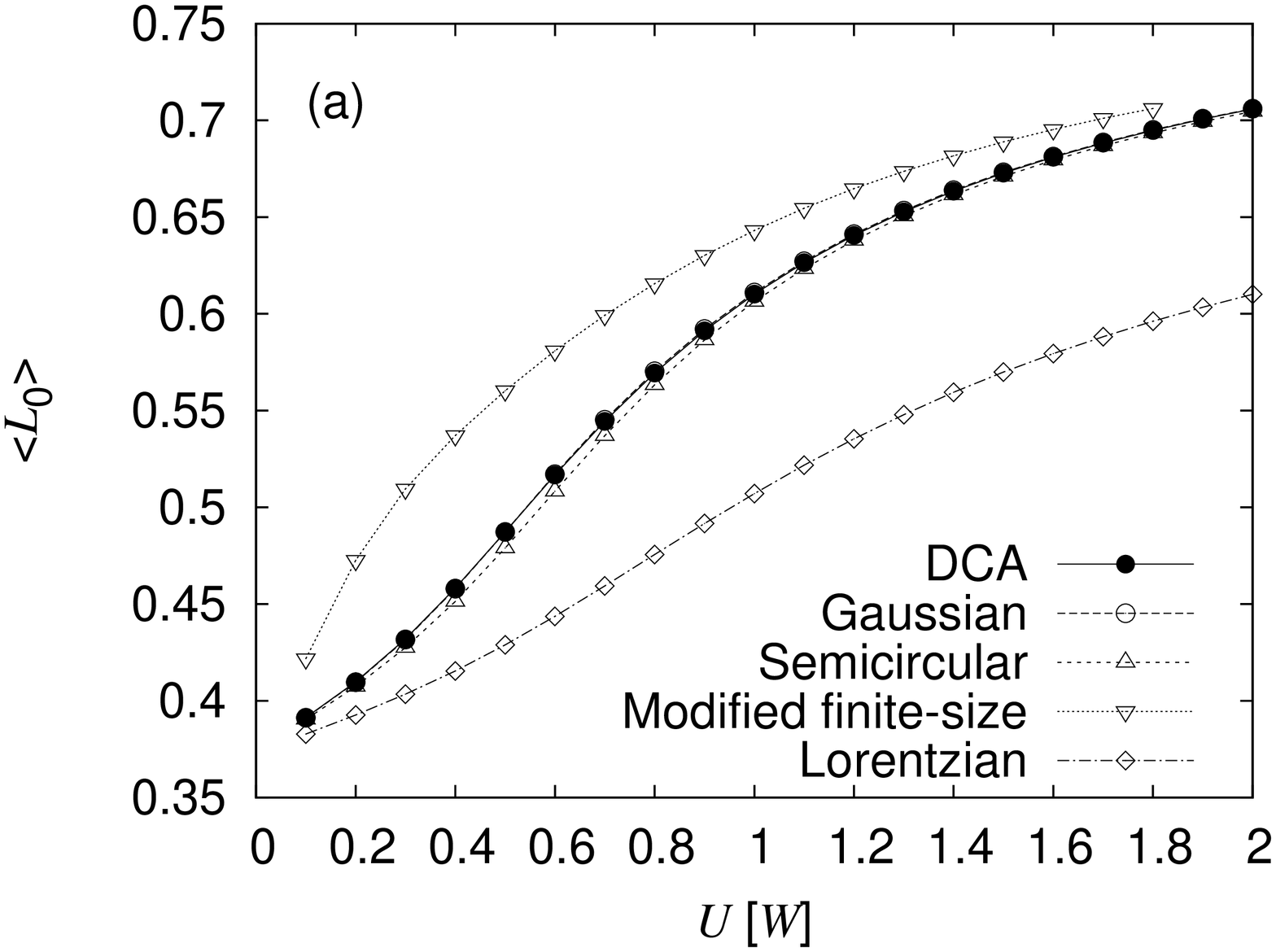}\\
\includegraphics[height=90mm,angle=270]{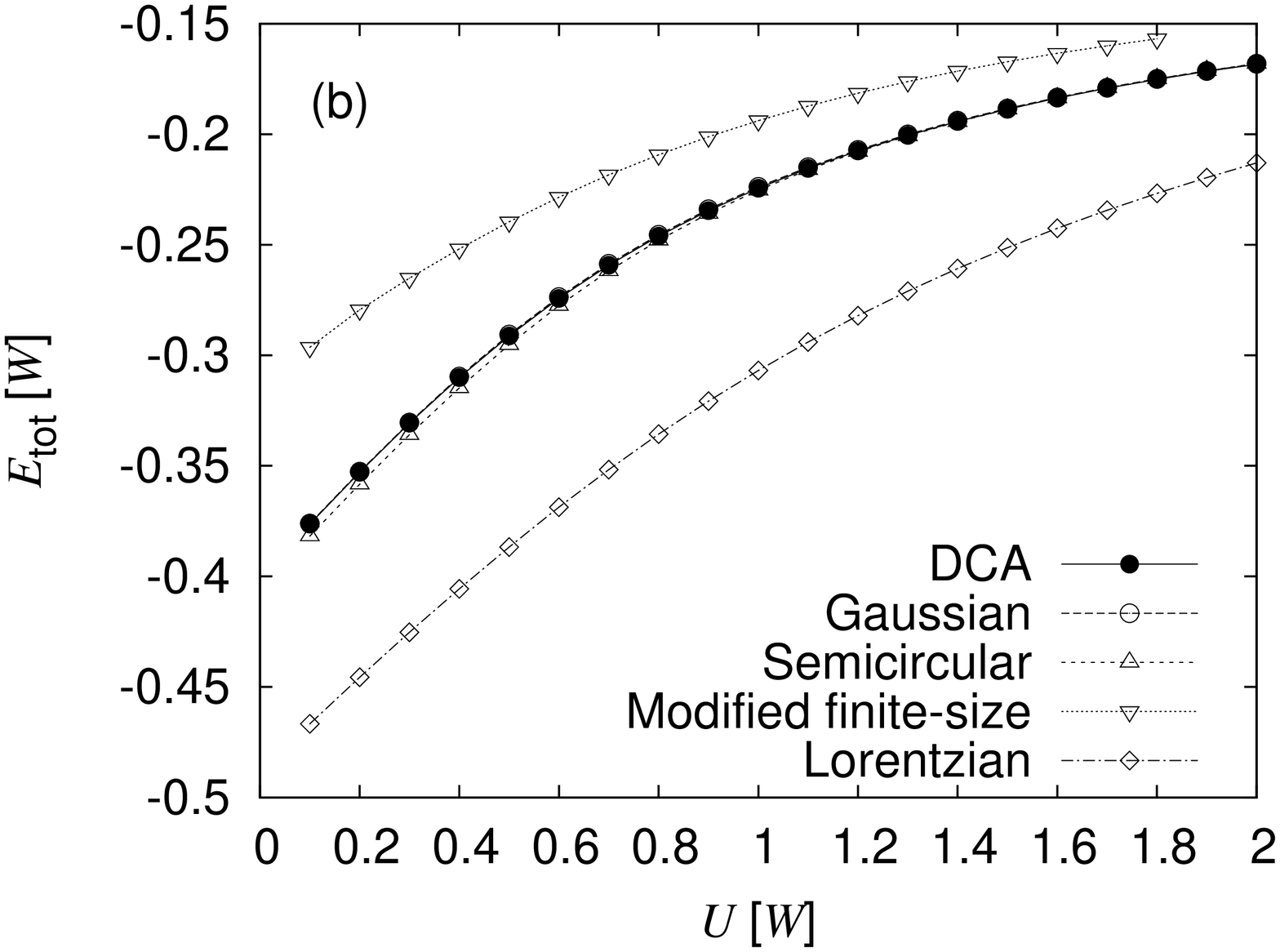}
\caption{Comparison between the DCA and the approximate DCA schemes
for thermodynamic quantities, using FLEX as a cluster solver and with
a cluster size $N_c=4$. Panel (a) shows the variation of the local
moment ($\langle L_0 \rangle$) with coupling, and panel (b) shows the
variation of the total energy, $E_{\mathrm{tot}}$, with coupling. The
temperature for these calculations is $T=0.04W$ and the density of
electrons per site, $n=1$. The Gaussian approximation to the DCA
self-consistency leads to results which are almost indistinguishable
from those calculated using the exact partial DOS (the open and filled circles lie on top of each other). The semi-circular
approximation to the partial DOS also leads to good results. It is
very important for the 1st and 2nd moments of the partial densities of
states to match, or there are significant inconsistencies, such as
those seen using the modified finite size approximation and Lorentzian
DOS. For methods with matching moments, agreement is almost exact for
$U\gtrsim W$.}
\label{figure:varu}
\end{figure}

\begin{figure}
\includegraphics[height=100mm,angle=270]{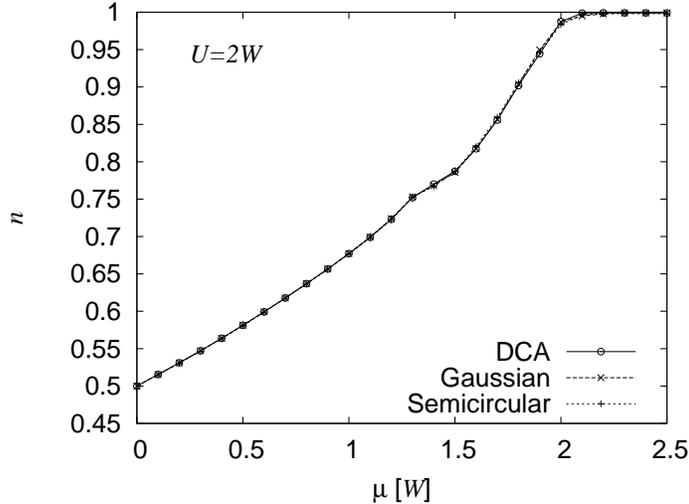}
\caption{Variation of filling with chemical potential showing the
performance of the scheme away from half filling. $T=0.04W$ and
$U=2W$. Agreement is good for filling less than $n=1.5$, where a kink
develops. For larger fillings, there are small discrepancies of the
order of a few percent. This is expected, since the Gaussian form of
the partial DOS has unphysical long tails, where the exact partial DOS has a
sharp cutoff.}
\label{figure:varmu}
\end{figure}

\begin{figure}
\includegraphics[height=100mm,angle=270]{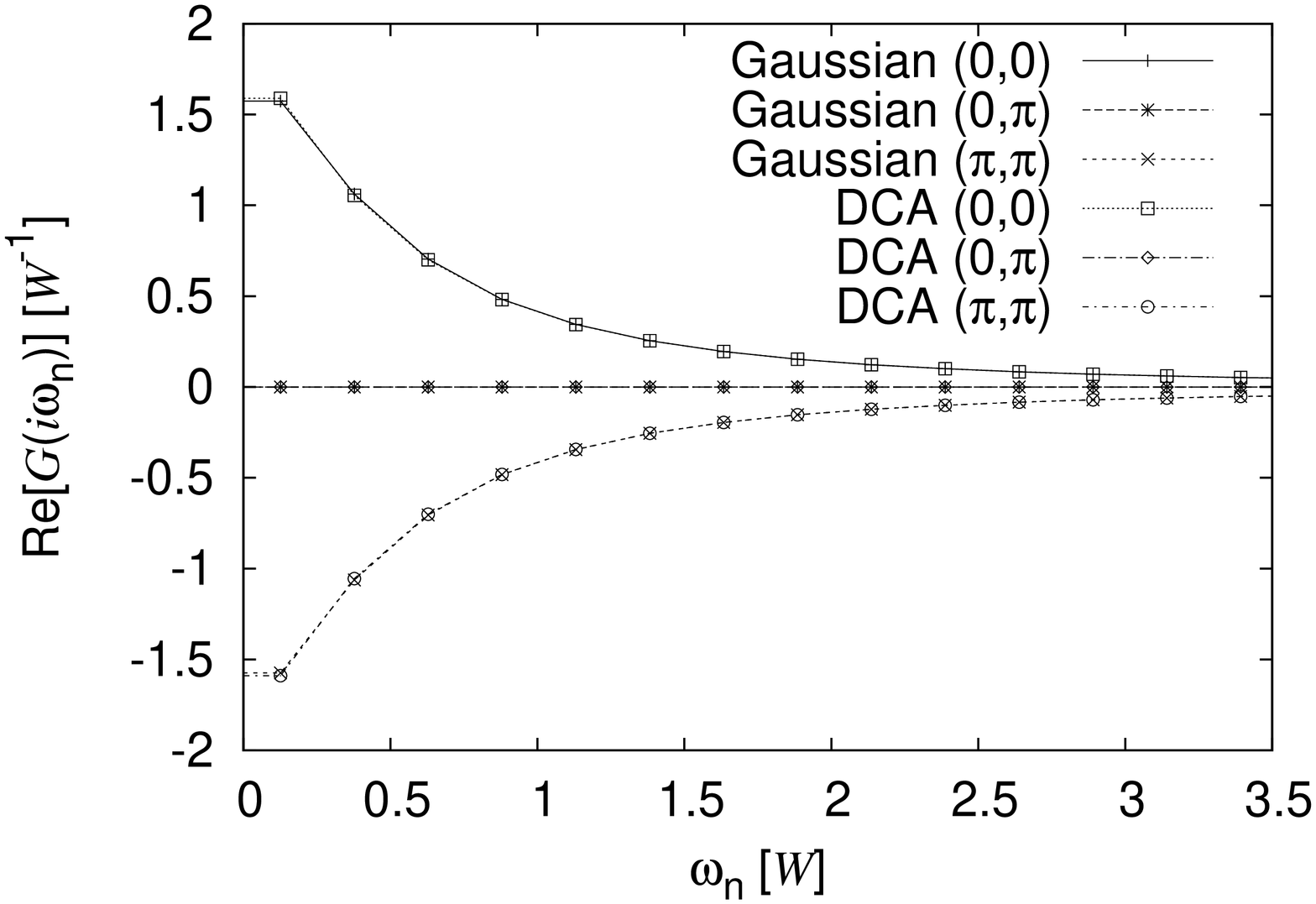}
\includegraphics[height=100mm,angle=270]{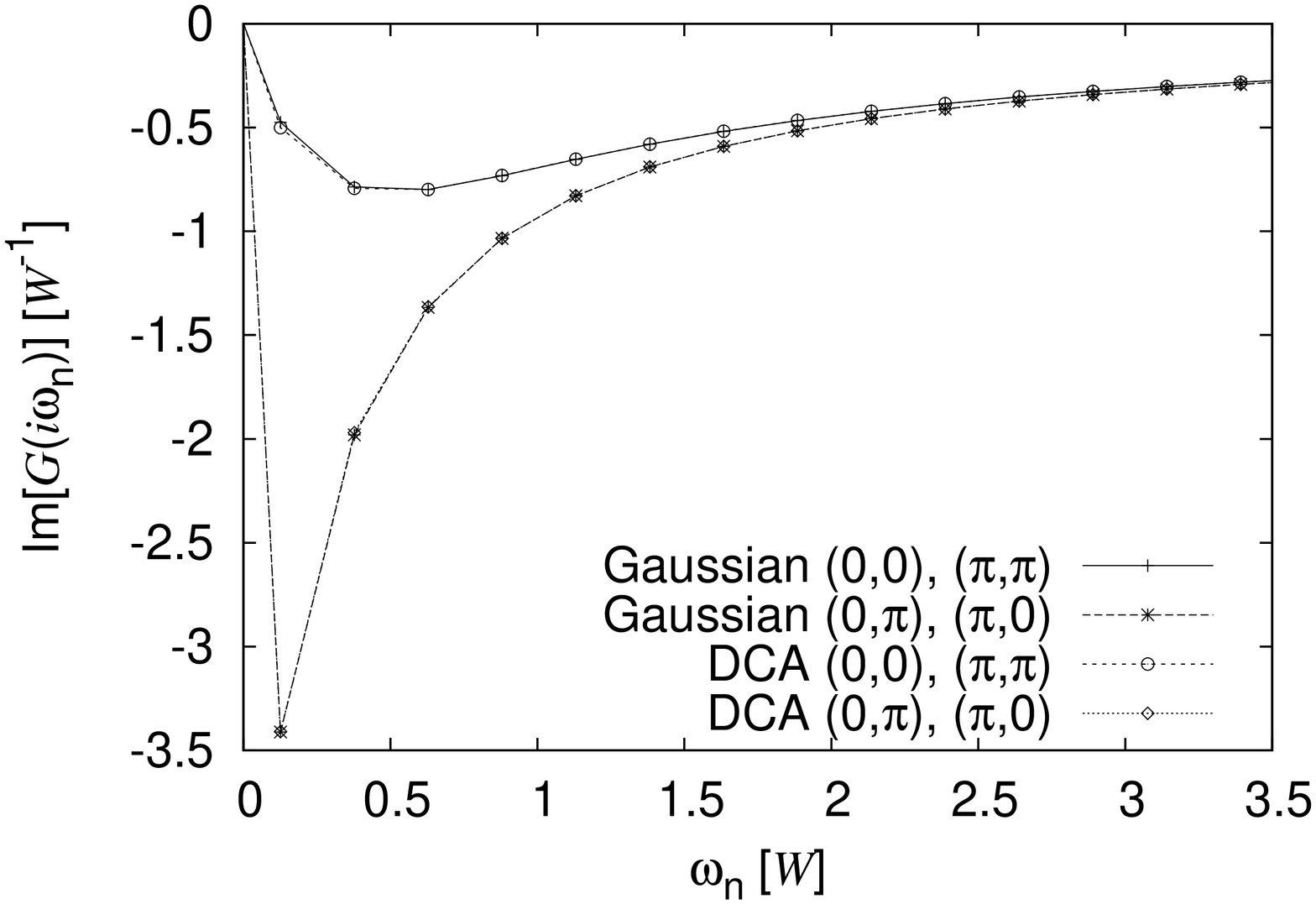}
\caption{Real and imaginary parts of the Green function on the Matsubara axis. Values are in agreement to within 1\%. $U=0.5W$, $N_{c}=4$ and $T=0.04W$ and the
system is half filled.}
\label{fig:realim}
\end{figure}

\begin{figure}
\includegraphics[height=100mm,angle=270]{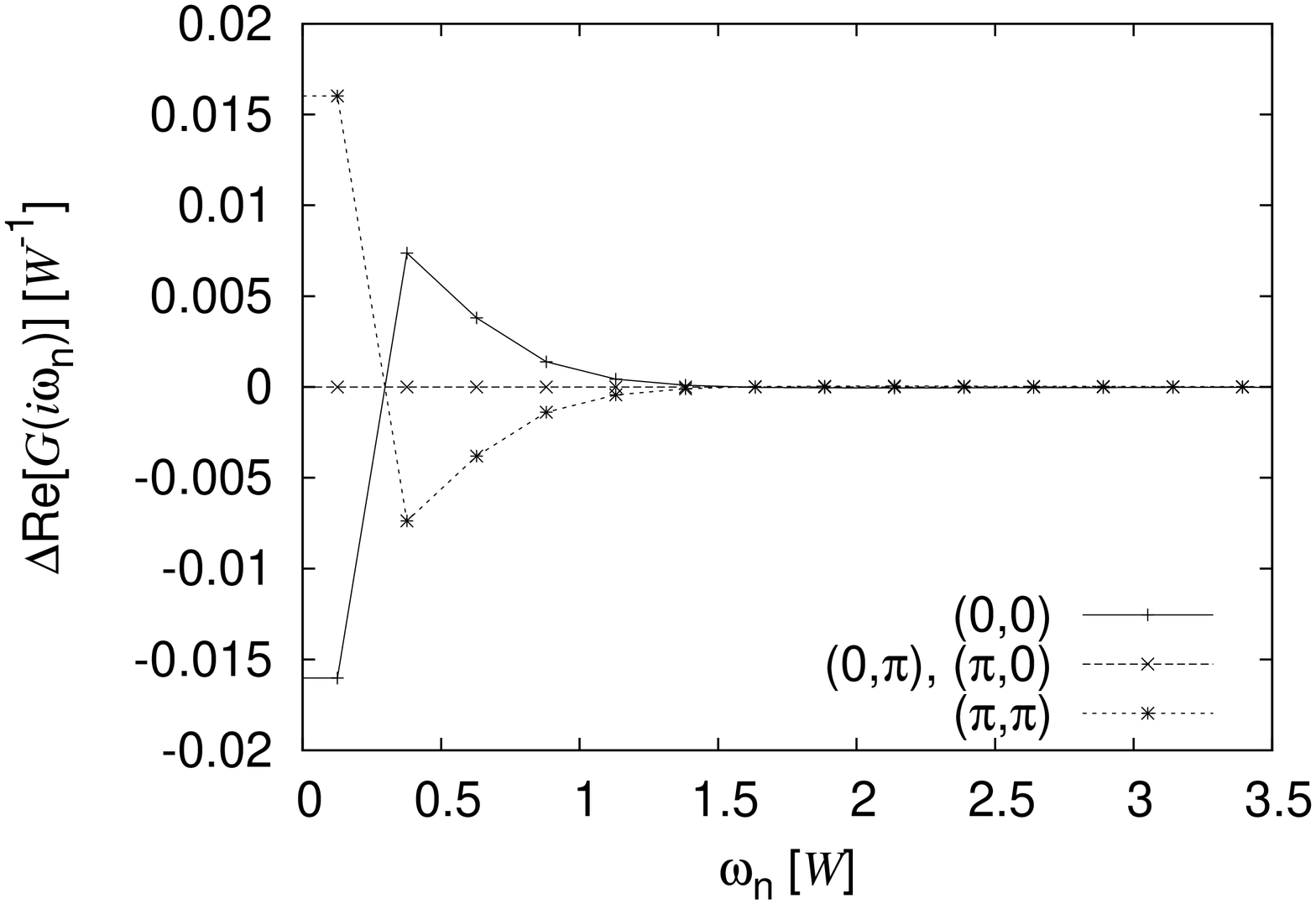}
\includegraphics[height=100mm,angle=270]{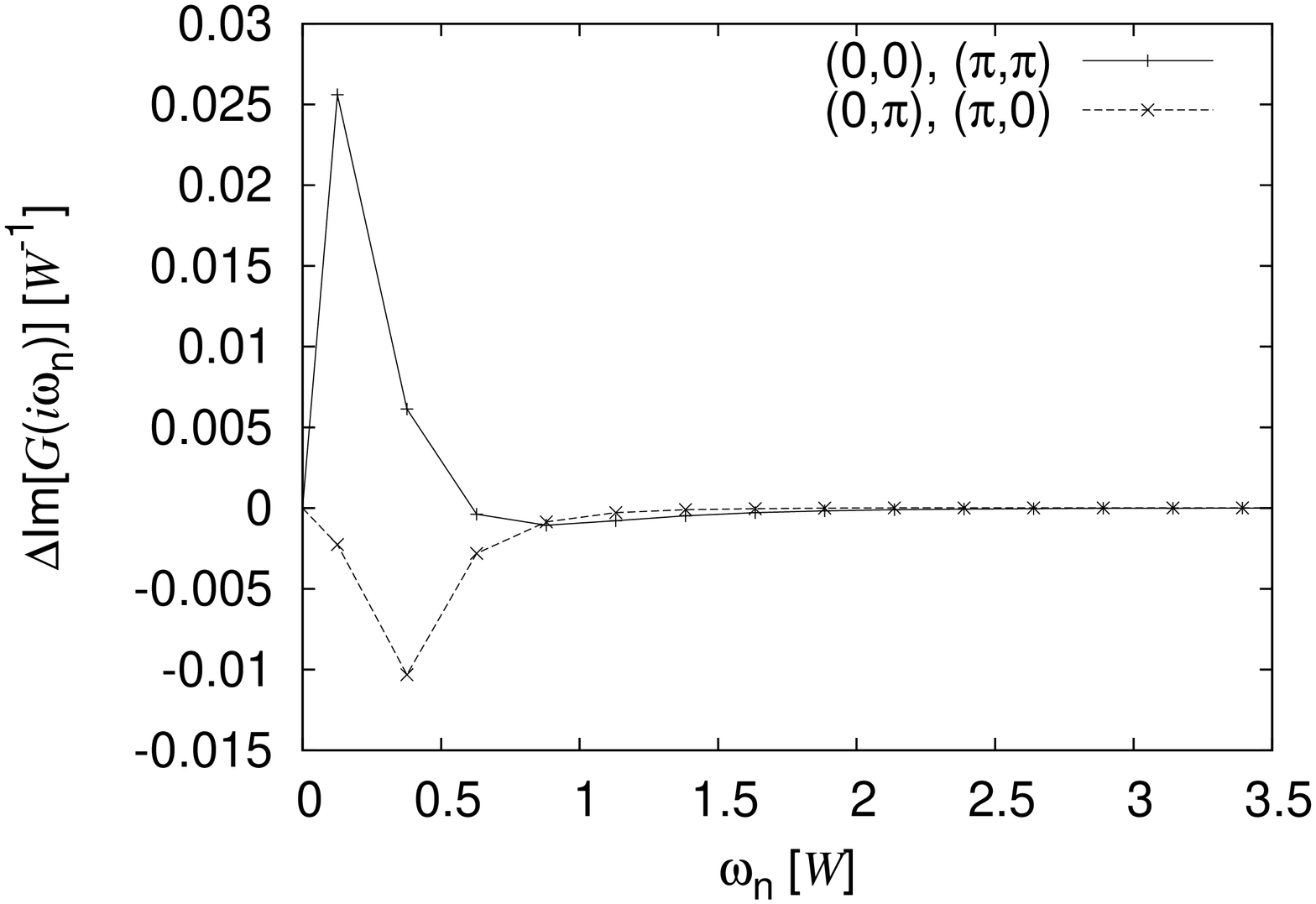}
\caption{Comparison of real and imaginary parts of the Green function on the Matsubara axis. Parameters are as Fig. \ref{fig:realim}.}
\label{fig:cmprealim}
\end{figure}

\begin{figure}
\includegraphics[height=100mm,angle=270]{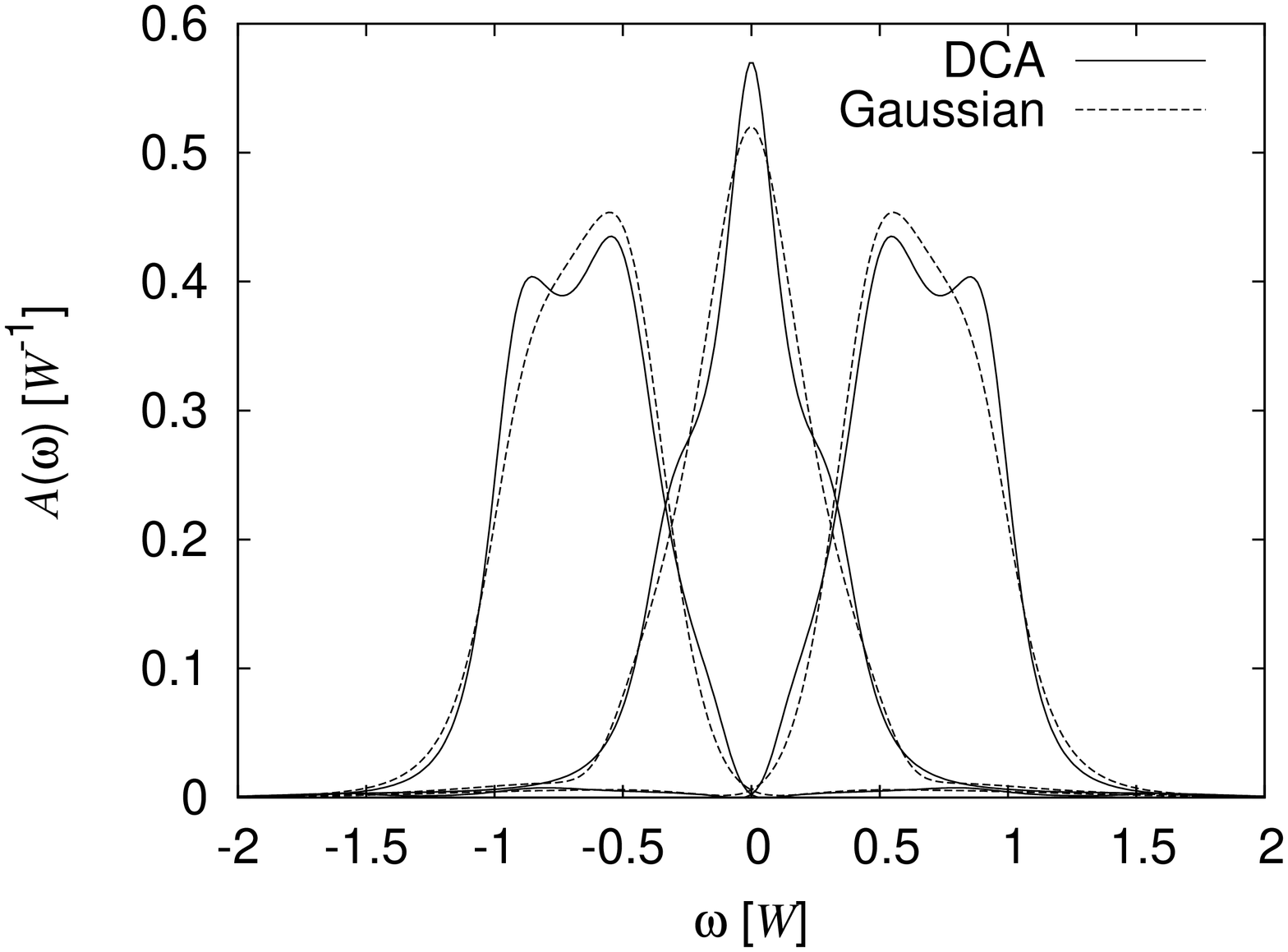}
\caption{Comparison of the density of states computed using a Pad\'e
approximant analytic continuation of order 800. While the detailed
structure of the interacting DOS varies slightly, the main features
are similar. This is in spite of the significant difference between
the exact and approximate forms of the non-interacting density of
states at the Fermi-surface. Parameters are as Fig. \ref{fig:realim}.}
\label{fig:cmppade}
\end{figure}

This section assesses the applicability of the approach described in
the last section. The accuracy of the results computed using the approximate self-consistent step is
considered as a function of the Hubbard $U$. Values of the Hubbard $U$ approaching the band-width will be considered. In
this paper, numerics are used to demonstrate the validity of the
scheme. However, the intention is that researchers will be able to use
approximate forms of the self-consistent condition to generate
analytic solutions of model systems in conjunction with appropriate
forms of the self energy.

In order to test the accuracy of a scheme which uses approximate
partial densities of states, the fluctuation-exchange (FLEX)
approximation is used to compute the self energy of the
cluster. Details of the FLEX approach and its use as a cluster solver
may be found elsewhere \cite{bickers1989a,aryanapour2002a}. FLEX has a
transparent physical interpretation, since it approximates the
self-energy by three subsets of diagrams, which represent spin-flips,
density fluctuations and pair fluctuations. Computations were carried
out at finite temperature using the Matsubara formalism. All
calculations were initialized with zero self-energy, and then the
Green function was calculated self-consistently until a convergence of
1 part in $10^{6}$ was reached. At this stage, observables were
calculated. Ideally a full quantum Monte-Carlo scheme would be used to
solve the cluster impurity problem, since FLEX can overestimate the
self-energy and could potentially be insensitive to the different
forms of the course graining schemes (for example the absence of van
Hove singularities). However for moderate couplings FLEX is accurate,
and it is expected that approximate schemes for the self-consistency
are most likely to fail when electrons are not localized (since
strongly localized electrons can be well described with a single site
impurity).

The total energy and local moment are plotted in figure
\ref{figure:varu} as coupling, $U$, is varied. The cluster size is
$N_C=4$, temperature $T=0.04W$ and electron density $n=1$. Results
computed using a number of different approximations to the partial DOS are
shown. The Gaussian approximation to the partial DOS leads to the most
accurate results. In fact, the difference between the curves computed
using the Gaussian and exact non-interacting DOS is not easily
distinguished by eye. Agreement is also found when the semi-circular
partial DOS is used to approximate the self-consistency, with an
accuracy of a few percent. Using the semi-circular DOS to approximate
the self consistency is more favorable to analytic approximations,
since the self-consistent equations also have a square root form and
can easily be inverted. At large $U$, the curves corresponding to the
Gaussian DOS, semi-circular DOS and the DCA scheme all converge. This
is expected, since at large $U$ at half-filling hopping is suppressed,
the system becomes local and the precise form of the non-interacting
DOS (which is related to the details of the intersite hopping and the
specific lattice) becomes irrelevant.

The matching of both moments is very important to achieve accurate
results. When only the first moment is matched, for example using the
modified finite size approximation, the errors on the moment and total
energy are significant. There is a large overestimation of the local
moment, and a huge underestimation of the total energy. Approximating the exact non-interacting density of states with a Lorentzian
generates the worst approximation. Presumably this is because the
second moment is ill defined, and the kinetic energy in the absence of
interaction is grossly overestimated. However, some of the qualitative features remain and it might be possible to
extract some useful physics even with such a crude approximation
\cite{georges1996a}.

The previous results were calculated for a half filled lattice,
corresponding to one of the interesting limits of the Hubbard
model. To determine the performance of the approximation away from
half-filling, figure \ref{figure:varmu} shows the variation of the
filling as the chemical potential, $\mu$, is changed. The temperature
is $T=0.04W$ and the coupling $U=2W$. Calculations are shown for the
full DCA self consistency scheme, and also for the Gaussian and
semi-circular approximations to the self-consistency. The change of
$n$ gives a rough measure of the variation in the renormalized DOS at
the Fermi energy for a particular value of the chemical potential, and
thus an indication of the performance of the scheme away from half
filling. For $n\neq 1$, it might be expected that the analytic schemes
have larger errors, since the approximate forms of the non-interacting
DOS do not have the correct form when the density of electrons is low
(for example the Gaussian DOS has unphysical high energy tails). In
fact, the agreement is good for filling less than $n=1.5$, where a
kink develops. For larger fillings, there are only small discrepancies
of the order of a fraction of a percent.

Since it is possible that cancellations of errors could lead to good
agreement of thermodynamic quantities, but may not give good results
when correlation functions are computed, Fig. \ref{fig:realim} shows
the real and imaginary components of the Green function on the
Matsubara axis. Only tiny differences can be seen between the forms of
the Green functions computed using the Gaussian approximated scheme and
the DCA. Here $U=0.5W$, $N_{c}=4$ and $T=0.04W$ and the system is half
filled. To clarify the quality of the approximation, the difference
between the values of the Green functions is shown in
Fig. \ref{fig:cmprealim}. For the parameters used here, the difference
is around 1\% for the smallest Matsubara frequencies, and the two
schemes converge for larger values of $\omega_n$. In the weak to
intermediate $U$ regime, any discrepancies between quantities computed
using the approximate and exact forms of the self-consistency
condition are expected to be most pronounced, and intermediate $U$ is
of most physical interest, so it is promising that the approximate
scheme works well for $U=0.5W$.

Finally, to demonstrate that the form of the interacting density of
states $A(\omega)=-{\rm Im}[G(\omega+i\eta)]/\pi$ can be computed to
reasonable accuracy using the alternative self consistency scheme, Pad\'{e} approximants are used to analytically continue the Green function
to the real axis \cite{vidbergserene} as shown in
Fig. \ref{fig:cmppade}. While the detailed structure of the
interacting DOS varies slightly, the main features are similar. This
is in spite of the significant difference between the exact and
approximate forms of the non-interacting density of states at the
Fermi-surface (for example, a factor of almost 2 difference can be
seen at the Fermi surface in Fig. \ref{figure:comparedos}). In
particular, a similar amount of spectral weight is shifted across the
Fermi surface in both cases. The density of states at the
Fermi-surface associated with the course grained zone centered about
the $(\pi,0)$ point is of similar size for both approximations, and the
band widening is similar.

\section{Extension to two-particle properties.}
\label{sec:extension}

In principle, the analytic schemes presented here could be used for
calculating two particle properties, such as the superconducting, charge
density wave and spin density wave susceptibilities. To examine the
potential for the computation of two particle properties, I use the
particle-hole susceptibility as an example. Following
Ref. \onlinecite{maier2001a}, the coarse-grained non-interacting
susceptibility can be computed using the following expression,
\begin{equation}
\bar{\chi}^{0}_{\sigma\sigma'}[(\qvec,i\nu_s);(\Kvec_i,i\omega_n)(\Kvec_j,i\omega_m)=\delta_{\sigma\sigma'}\delta_{ij}\delta_{nm}\sum_{\tilde{\kvec}}G_{\sigma}(\Kvec_i+\tilde{\kvec},i\omega_n)G_{\sigma}(\Kvec_i+\tilde{\kvec}+\qvec,i\omega_n + i\nu_{s}) \, ,
\end{equation}
where $\tilde{\kvec}$ is a vector that lies within a sub-Brillouin zone. It is possible to simplify this expression for use with the approximate scheme discussed in this article in two special cases; when $\qvec=\zerovec=(0,0,\cdots)$ and when $\qvec=\pivec=(\pi,\pi,\cdots)$. First, if $\qvec=\zerovec$ then the propagator becomes,
\begin{eqnarray}
& & \bar{\chi}^{0}_{\sigma\sigma'}[(\zerovec,i\nu_s);(\Kvec_i,i\omega_n)(\Kvec_j,i\omega_m) = \delta_{\sigma\sigma'}\delta_{ij}\delta_{nm}\times \\ & & \sum_{\tilde{\kvec}}\frac{1}{i\omega_n+\mu-\epsilon(\kvec)-\Sigma(i\omega_n,\Kvec_i+\tilde{\kvec})}\frac{1}{i\omega_n+i\nu_{s}+\mu-\epsilon(\kvec)-\Sigma(i\omega_n+i\nu_{s},\Kvec_i+\tilde{\kvec})} \, .\nonumber
\end{eqnarray}
Since in the dynamical cluster approximation $\Sigma$ is constant within each sub-zone, this simplifies to,
\begin{eqnarray}
\label{eqn:qzero}
& &\bar{\chi}^{0}_{\sigma\sigma'}[(\zerovec,i\nu_s);(\Kvec_i,i\omega_n)(\Kvec_j,i\omega_m)=\delta_{\sigma\sigma'}\delta_{ij}\delta_{nm}\times \\ & &\int\frac{D_{\Kvec_i}(\epsilon)}{(i\omega_n+\mu-\epsilon-\Sigma(i\omega_n,\Kvec_i))(i\omega_n+i\nu_s+\mu-\epsilon-\Sigma(i\omega_n+i\nu_s,\Kvec_i))} \, , \nonumber
\end{eqnarray}
so that the form of the partial density of states consistent with the approximation to the self-consistent equation can be used. Equation \ref{eqn:qzero} is appropriate for all forms of lattice.

Simplifications can also be made in the more useful case where $\qvec=\pivec$ if there is particle-hole symmetry, for example if the lattice is square (or hypercubic) and only has near-neighbor hopping. Then $\epsilon(\kvec)=-t\sum_{n=1}^{d}\cos(k)$, so $\epsilon(\kvec+\pivec)=-\epsilon(\kvec)$, and the susceptibility simplifies to
\begin{eqnarray}
& & \bar{\chi}^{0}_{\sigma\sigma'}[(\pivec,i\nu_s);(\Kvec_i,i\omega_n)(\Kvec_j,i\omega_m)= \\ & & \delta_{\sigma\sigma'}\delta_{ij}\delta_{nm}\int\frac{D_{\Kvec_i}(\epsilon)}{(i\omega_n+\mu-\epsilon-\Sigma(i\omega_n,\Kvec_i))(i\omega_n+i\nu_s+\mu+\epsilon-\Sigma(i\omega_n+i\nu_s,\Kvec_i+\pivec))}.\nonumber
\label{eqn:qpi}
\end{eqnarray}

The two particle cluster susceptibility $\chi_{c,\sigma\sigma'}[(\qvec,i\nu_s);(\Kvec_i,i\omega_n)(\Kvec_j,i\omega_m)]$ is computed in a manner which is specific to the cluster solver that is used, and does not involve any coarse graining. Once the cluster properties have been computed, the coarse grained lattice susceptibility can be computed using,
\begin{equation}
\bar{\chi}^{-1}=\chi_{c}^{-1}-[\chi_{c}^{0}]^{-1}+[\bar{\chi}^{0}]^{-1} \, ,
\end{equation}
from which the spin and charge susceptibilities can be computed \cite{maier2001a}. Calculation of $\chi_{c}$ is quite involved in the FLEX scheme and is not discussed here.

\section{Multiple bands}
\label{section:multiband}

\begin{figure}
\includegraphics[width=100mm]{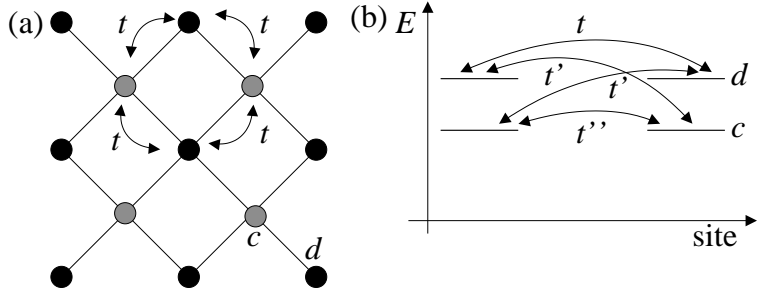}
\caption{Possible hopping scenarios for multiband models that can be treated using the approximate self-consistent scheme (a) two bands arise from sites of different energies in a lattice with a basis of two atoms (b) two bands arise from hopping between electrons in different orbitals.}
\label{fig:multibandschematic}
\end{figure}

I complete this article with a discussion of the applicability of the
approximate scheme to multiple bands. I consider only two bands here,
but similar arguments will be applicable when there are more than two
bands. A very generic two-band Hamiltonian has the form,
\begin{equation}
H_{2}=\sum_{\kvec}\epsilon_{\kvec}\cdagg_{\kvec}c_{\kvec}+\zeta_{\kvec}\ddagg_{\kvec}d_{\kvec}+\gamma_{\kvec}(\cdagg_{\kvec}d_{\kvec}+\ddagg_{\kvec}c_{\kvec}) (+ \, H_{2, \rm int}) \, ,
\end{equation}
where $\cdagg$ and $\ddagg$ create electrons of distinct types (either
on different atoms of a two-atom basis, or in different orbitals of a
single atom). $\epsilon_{\kvec}$ is the electronic dispersion
generated from hopping of electrons of type $c$, and $\zeta_{\kvec}$
corresponds to electrons of type $d$. The dispersions
$\epsilon_{\kvec}$ and $\zeta_{\kvec}$ typically correspond to hopping
between orbitals of the same type on different sites as shown
schematically as hops of amplitude $t$ and $t''$ in
Fig. \ref{fig:multibandschematic}(b). $\gamma_{\kvec}$ is the
dispersion of electrons that hop between different types of atom such as
might be found in crystals with a two-atom basis (shown schematically
in Fig. \ref{fig:multibandschematic}(a)). Dispersion of type
$\gamma_{\kvec}$ can also originate from hopping between orbitals of
different types on neighboring sites, as shown schematically as hops
of amplitude $t'$ in Fig. \ref{fig:multibandschematic}(b). $H_{2, \rm
int}$ contains all the interaction terms between electrons of types
$c$ and $d$ within the cell.

Diagonalization of the quadratic terms using a standard Bogoliubov
transformation yields,
\begin{equation}
H_{2} = E^{(\alpha)}_{\kvec}\alphadagg_{\kvec}\alpha_{\kvec}+E^{(\beta)}_{\kvec}\betadagg_{\kvec}\beta_{\kvec} \, ( + H_{2,\rm int} ) \, ,
\end{equation}
where,
\begin{eqnarray}
E^{(\alpha),(\beta)}& =& \left[\frac{\epsilon_{\kvec}}{2}\left(1\pm\frac{\epsilon_{\kvec}-\zeta_{\kvec}}{\sqrt{4\gamma_{\kvec}^2+(\epsilon_{\kvec}
-\zeta_{\kvec})^2}}\right)+\frac{\zeta_{\kvec}}{2}\left(1\mp\frac{\epsilon_{\kvec}-\zeta_{\kvec}}{\sqrt{4\gamma_{\kvec}^2+(\epsilon_{\kvec}-\zeta_{\kvec})^2}}\right) \right.\nonumber\\
& &\hspace{30mm} \left.\pm \frac{2\gamma_{\kvec}^2}{\sqrt{4\gamma_{\kvec}^2+(\epsilon_{\kvec}-\zeta_{\kvec})^2}}\right] \, ,
\end{eqnarray}
$\cdagg_{\kvec} =
u_{\kvec}\alphadagg_{\kvec}+v_{\kvec}\betadagg_{\kvec}$,
$\ddagg_{\kvec} =
-v_{\kvec}\alphadagg_{\kvec}+u_{\kvec}\betadagg_{\kvec}$, 
$u_{\kvec}^2=(1+A_{\kvec})/2$, $v_{\kvec}^2=(1-A_{\kvec})/2$, $u_{\kvec}v_{\kvec}=\sqrt{1-A_{\kvec}^2}/2$ and
$A_{\kvec}=(\epsilon_{\kvec}-\zeta_{\kvec})/\sqrt{4\gamma_{\kvec}^2+(\epsilon_{\kvec}-\zeta_{\kvec})^2}$. The non-interacting Matsubara Green functions for $\tau>0$ form a $2\times 2$ matrix with the elements, $G^{(0)}_{cc}(\kvec,\tau)=-\langle c_{\kvec}(\tau)\cdagg_{\kvec}(0)\rangle$ and $G^{(0)}_{cd}(\kvec,\tau)=-\langle c_{\kvec}(\tau)\ddagg_{\kvec}(0)\rangle$ etc. which can be computed by substituting the Bogoliubov transformation (since $\alpha$ and $\beta$ excitations have a Fermi-Dirac distribution). After Fourier transformation,
\begin{equation}
G^{(0)}_{cc}(\kvec,i\omega_n) = \left[\frac{u_{\kvec}^2}{i\omega_n-E^{(\alpha)}_{\kvec}}+\frac{v_{\kvec}^2}{i\omega_n-E^{(\beta)}_{\kvec}}\right]
\end{equation}
and
\begin{equation}
G^{(0)}_{cd}(\kvec,i\omega_n)= \frac{u_{\kvec}v_{\kvec}\left[E^{(\alpha)}_{\kvec}-E^{(\beta)}_{\kvec}\right]}{(i\omega_n-E^{(\alpha)}_{\kvec})(i\omega_n-E^{(\beta)}_{\kvec})}.
\end{equation}
%

When interaction is switched on, the coarse grained Green function is,
\begin{equation}
\boldsymbol{G} = \sum_{\tilde{\kvec}} \boldsymbol{G}_{0} \left[ 1+\boldsymbol{G}_0 \boldsymbol{\Sigma} \right]^{-1}
\end{equation}
where the self-energy, $\boldsymbol{\Sigma}$, is also a $2\times 2$ matrix. As in the one-band case, the self-energy is constant within each sub-zone.

In contrast to the one-band case, the sum on $\tilde{\kvec}$ cannot always be
replaced by a density of states. This is because $\epsilon_{\kvec}$,
$\zeta_{\kvec}$ and $\gamma_{\kvec}$ do not normally have the same
form. Under certain conditions it is possible to make the replacement
and therefore use the analytic forms of the self-consistency
considered in this article. In the first case, bands with
dispersion $\gamma_{\kvec}$ are formed via nearest-neighbor hops in a
system with a basis of two atoms, and $\epsilon_{\kvec}$ and
$\zeta_{\kvec}$ are flat, representing the energies of the atomic
sites. Then the sum over momenta is replaced by an integral over the
density of states corresponding to $\gamma_{\kvec}$, which can then be
replaced using one of the simplified analytic forms considered in
section \ref{section:dca}. I note that two-band systems of this type
have been studied using DMFT \cite{georges1993a}.

The sum over momenta can also be replaced with an integral over a
density of states when $\epsilon_{\kvec}\propto\zeta_{\kvec}$,
$\epsilon_{\kvec}\propto\gamma_{\kvec}$ or
$\zeta_{\kvec}\propto\gamma_{\kvec}$ (up to an offset) so long as any
other dispersions are flat. This is possible for electrons moving
between atoms with several orbitals by nearest neighbor hopping only
(the hopping integrals can be different) as shown schematically in
Fig. \ref{fig:multibandschematic}(b). Then the density of states
corresponding to only one of the dispersions can be used, since any
dispersion can be represented in terms of the other dispersion. In
general systems with next-nearest neighbor hopping are excluded from
this argument unless there are good grounds for expecting the ratio of
nearest to next-nearest neighbor hopping integrals to be identical for
every band. There may be other more complicated non-linear maps
between dispersions (e.g. $\epsilon_{\kvec}=F(\gamma_{\kvec})$ where
$F$ is an arbitrary function) so that the same trick can be applied,
however the mapping must be $\kvec$ independent so such mappings will
also be special cases.

\section{Summary}
\label{section:summary}

This paper introduced approximate schemes for the self-consistent step
of the dynamical cluster approximation. In order to simplify the
self-consistent step, the partial density of states used to calculate
the Green function was replaced with a simple analytic form. The limitations
and errors associated with the approach were tested using the FLEX
approximation for the self energy. The approximate schemes were
demonstrated to work well for intermediate coupling over a range of
fillings.

The best approximate result comes from replacing the exact
non-interacting DOS with a Gaussian. Approximating using a
semi-circular partial DOS also leads to agreement of thermodynamic
properties to within a few percent of the exact result. The
semi-circular approximation leads to self-consistent equations with a
much simpler form, which is promising for analytic work. Finally,
forms such a Lorentzian and a modified finite size scheme, where only
the 1st moment is matched, lead to quantitatively different results,
although some of the qualitative physics remains. These schemes lead
to even simpler forms for the self-consistency. Examination of the
Green function and carrying out an analytic continuation demonstrate
that the scheme can also compute dynamic properties to reasonable
accuracy. I add the caveat that some properties of the 2D Hubbard
model may show features consistent with the influence of van Hove
singularities \cite{gull2009a}, and such singularities are not present
in the simple approximations with small cluster sizes (although they
will emerge in very large clusters). In 3D, there are no divergences
in the non-interacting density of states and spatial fluctuations are
less relevant, so the schemes may be most effective for such problems.

Given that the analytic forms for the self-consistency introduced here
work effectively, the formalism could provide a good starting point
for the calculation of analytic phase diagrams and other properties in
low dimensional systems where non-local fluctuations are essential
(for example unconventional superconductivity such as $d$, $p$ and
extended-$s$ wave which cannot be examined with the DMFT). It is hoped
that this article will stimulate such studies.

\section{acknowledgements}

I acknowledge support from EPSRC grant no. EP/H015655/1.

\vspace{5mm}


\end{document}